\documentclass[preprint]{emulateapj} 

\usepackage{natbib}
\usepackage{hyperref}
\hypersetup{colorlinks,citecolor=Blue,linkcolor=Red,urlcolor=Blue}
\usepackage{amsmath}
\usepackage{empheq}
\usepackage[usenames,dvipsnames]{color}

\allowdisplaybreaks 
\hypersetup{colorlinks,citecolor=Blue,linkcolor=Red,urlcolor=Blue}
\usepackage[usenames,dvipsnames]{color}
\usepackage{amsmath,amsfonts,amssymb}
\usepackage{graphicx}
\usepackage{wasysym}
\usepackage{float}
\usepackage{url}
\newcommand{\appropto}{\mathrel{\vcenter{\offinterlineskip\halign{\hfil$##$\cr\propto\cr\noalign{\kern2pt}\sim\cr\noalign{\kern-2pt}}}}}

\begin{document}

\shorttitle{Updated Masses for the TRAPPIST-1 Planets}

\title{Updated Masses for the TRAPPIST-1 Planets}  
\author{Songhu Wang$^{1}$, Dong-Hong Wu$^{2}$, Thomas Barclay$^{3,4}$, and Gregory P. Laughlin$^{1}$} 
\affil{$^{1}$Department of Astronomy, Yale University, New Haven, CT 06511, USA}
\affil{$^{2}$School of Astronomy and Space Science and Key Laboratory of Modern Astronomy and Astrophysics in Ministry of Education, Nanjing University, Nanjing 210093, China} 
\affil{$^{3}$NASA Goddard Space Flight Center, 8800 Greenbelt Rd, Greenbelt, MD 20771, USA} 
\affil{$^{4}$University of Maryland, Baltimore County, 1000 Hilltop Cir, Baltimore, MD 21250, USA}
\email{song-hu.wang@yale.edu}

\begin{abstract} 
The newly detected TRAPPIST-1 system, with seven low-mass, roughly Earth-sized planets transiting a nearby ultra-cool dwarf, is one of the most important exoplanet discoveries to date. The short baseline of the available discovery observations, however, means that the planetary masses (obtained through measurement of transit timing variations of the planets of the system) are not yet well constrained. The masses reported in the discovery paper were derived using a combination of photometric timing measurements obtained from the ground and from the Spitzer spacecraft, and have uncertainties ranging from 30\% to nearly 100\%, with the mass of the outermost, $P=18.8\,{\rm d}$, planet h remaining unmeasured. Here, we present an analysis that supplements the timing measurements of the discovery paper with 73.6 days of photometry obtained by the K2 Mission. Our analysis refines the orbital parameters for all of the planets in the system. We substantially improve the upper bounds on eccentricity for inner six planets (finding $e<0.02$ for inner six known members of the system), and we derive masses of $0.79\pm 0.27 \,M_{\oplus}$, $1.63\pm 0.63\,M_{\oplus}$, $0.33\pm 0.15\,M_{\oplus}$,  $0.24^{+0.56}_{-0.24}\,M_{\oplus}$, $0.36\pm 0.12\,M_{\oplus}$, $0.566\pm 0.038\,M_{\oplus}$, and $0.086\pm 0.084\,M_{\oplus}$ for planets b, c, d, e, f, g, and h, respectively.
\end{abstract} 

\maketitle
\section{Introduction}
The planetary systems of the lowest-mass stars have attracted the attentions of the astronomical community. For worlds of given mass or radius, the prospects for detection via either transit photometry or Doppler velocimetry are enhanced for red dwarf primaries. Furthermore, it is clear that low-mass stars very frequently serve as planet hosts. Radial velocity programs have indicated that the average M-dwarf is accompanied by one or more detectable planets \citep{tuomi2014}, whereas transit surveys, notably the \textit{Kepler} Mission, have provided evidence for $2.5\pm0.2$ low-mass planets per M-dwarf \citep{dressing2015}.

Furthermore, the transit depths for super-Earths and terrestrial-sized planets orbiting red dwarf primaries are large enough to permit their detection from the ground using small telescopes. Over the past decade, notable transiting red dwarf planetary companions include the Neptune-mass planet orbiting Gliese 436 \citep{gillon2007}, the super-Earth Gliese 1214b \citep{charbonneau2009}, and most dramatically, the seven terrestrial-sized planets orbiting the 12.1 pc-distant M8V dwarf 2MASS J23062928-0502285, now famously known as TRAPPIST-1.

The first detections associated with the TRAPPIST-1 system were announced by \citet{gillon2016}, who identified three transiting Earth-sized planets on the basis of ground-based photometric monitoring. These detections prompted further, high-priority scrutiny of the star, including extensive coverage with the Spitzer Space Telescope. These additional observations allowed \citet{Gillon2017} to provide a much-expanded assessment of the system. Seven terrestrial-sized planets, with orbital periods $P=$1.5, 2.4, 4.0, 6.1, 9.2, 12.4, and 18.8 days \citep{luger2017} are now known. Remarkably -- despite the very small planetary masses -- the proximity of the orbits to low-order mean motion resonances induces substantial transit timing variations (TTVs). Dynamical modeling of these variations allow the individual planetary masses to be inferred. \citet{Gillon2017}'s TTV analysis extracts planetary masses of order $M_{\rm p}\sim M_{\oplus}$ (to $\sim$2$\sigma$ precision) for all of the inner six planets, albeit with uncertainties in the mass determinations that range from 30\% to nearly 100\%.

Given the various conjectures that Earth-like environments could potentially exist on one or more of the TRAPPIST-1 planets, there is intense interest in improving both the mass determinations and the orbital parameters of the planets through the measurement of additional transit timing variations. With these refinements, the basic physical characteristics of the planets can be better assessed, and the resonant dynamical relationship between the planets (which may give direct insights into their formation process and the prospects for the ultra-long term stability of the system) can be elucidated.

Through a fortuitous coincidence, TRAPPIST-1 was very recently observed by NASA's \textit{Kepler} spacecraft as part of the K2 Mission's Campaign 12 \citep{Howell2014}. The star, and its surrounding 110-square degree field was monitored almost continuously, starting on December 15, 2016 and concluding on March 4, 2017. The resulting photometric time series captured new transits for all seven known planets in the system \citep{luger2017}, and the quality of the data permit accurate transit timing measurements. In this short paper, we present an analysis of this data, in combination with the previously published observations, that focuses on improving the measurements of the masses and the orbits of the planets in the system. 

We proceed in the following manner: In \S2, we give an overview of the observational data and present our reduction that generates a 73.6-day light curve for the star. In \S3, we analyze the light curve and measure the individual times of occurrence for the embedded transits. In \S4, we use a model that accounts for dynamical planet-planet interactions to generate an updated extraction of the system parameters, and in \S5, we give a brief assessment of the ways that the new model departs from the old, and outline the prospects for immediate future progress in elucidating the nature this landmark planetary system. 

\section{Observation and Data Reduction}
TRAPPIST-1 was observed by the Kepler spacecraft as part of Campaign 12 of NASA's ongoing K2 mission \citep{Howell2014}. This observing campaign, which was centered on a field in the constellation Aquarius, began on Dec 15, 2016, and lasted for 78.9 days, ending on March 4, 2017. There was a 5.3 day gap in data collection starting on Feb 1 as a result of the spacecraft entering safe mode. As a consequence, a total of 73.6 days of K2 data were collected.

These data were downlinked to Earth and rapidly made publicly available at the Mikulski Archive for Space Telescopes (MAST) in a raw file format known as a cadence pixel file. The unprocessed pixel data containing TRAPPIST-1 were extracted from the cadence pixel files, converted to a file format similar to the standard Kepler target pixel file data \citep{archivemanual} using the kadenza software package \citep{Barentsen2017a}, and made available for download \citep{Barentsen2017b}. We adjusted the time-stamps on these files to correct for the spacecraft's Barycentric velocity.

\begin{figure}
\vspace{0cm}\hspace{0cm}
\includegraphics[scale=0.34]{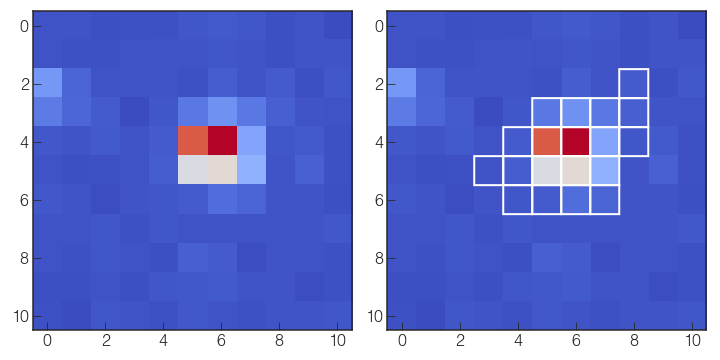}
\caption{The pixels used to create the light curve of TRAPPIST-1. The left panel shows an image at a single epoch and the right panel shows the same image with our adopted pixel mask overlaid.
\label{fig1}}
\end{figure}

For our analysis we use the short cadence (1-min time-sampling) data selected for catalog number EPIC 200164267. This provides us with an $11\times11$ pixel image at every time step in the sequence of observations. To create a light curve, we began by removing background and smear though subtracting the median value along each column, independently for every image. While this is significantly less rigorous than the image calibration performed in the Kepler pipeline \citep{Quintana2010}, for a faint source like TRAPPIST-1, the impact of this simplification is minimal. We empirically generate a photometric aperture by calculating the median image over all observations and including all pixels contiguous with the central pixel whose counts are 2.5 median absolute deviations (MAD) above the median pixel count. In Figure~\ref{fig1} we show the photometric aperture we selected. We summed all the pixel counts in the aperture at every time-step, creating a light curve. Additionally, we measured the position of the star at each observation by calculating the moments of the image.

\begin{figure*}
\vspace{0cm}\hspace{0cm}
\includegraphics[scale=0.33]{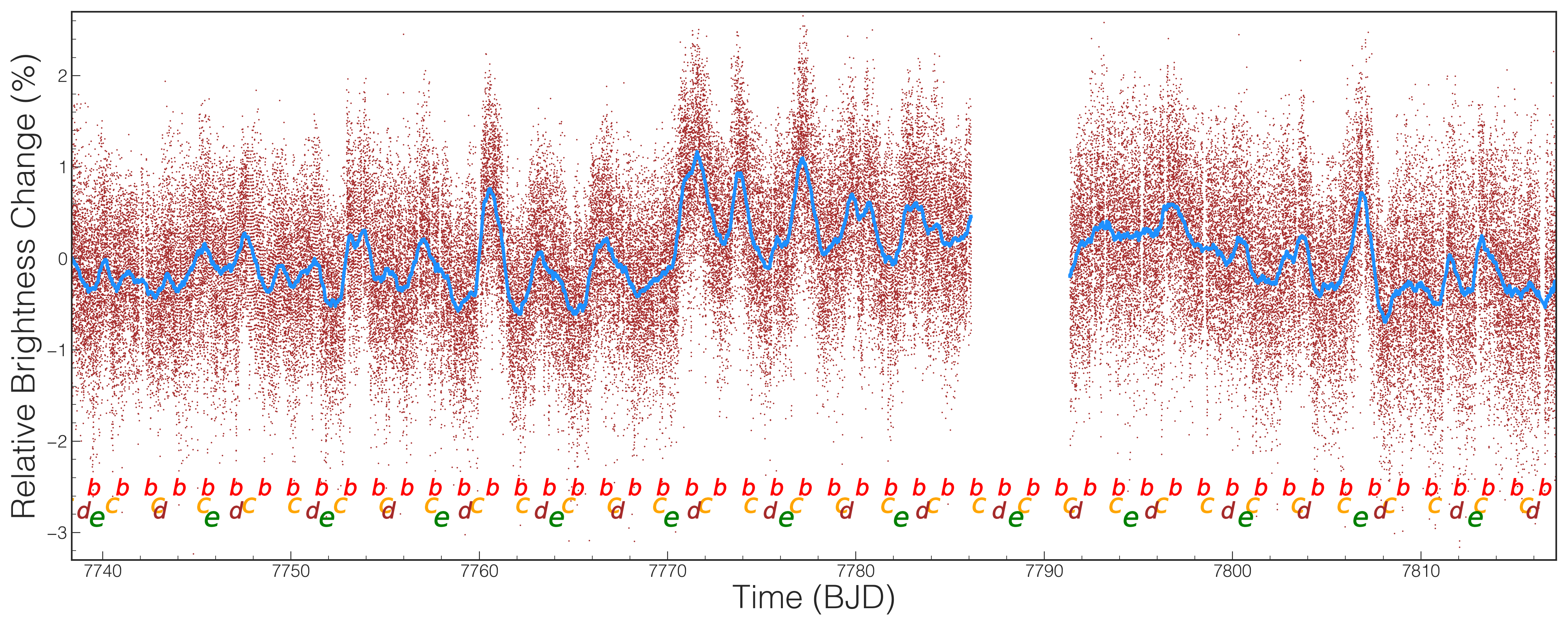}
\caption{The complete short-cadence light curve of TRAPPIST-1 is shown as red points. Instrumental signals have already been removed from these data. The blue curve is a running median, and shows a characteristic modulation due to spots on a rotating star with a period of approximately 3 days. The transit time for each planet is marked by the corresponding  planet letter.
\label{fig2}}
\end{figure*}

Our light curve contains two sources of variability that are not related to the planetary transits: (1) instrumental signals from the spacecraft and (2) variability intrinsic to the star. The primary source of instrumental signals is due to roll motion. The Kepler spacecraft can no longer maintain long-duration fine pointing with only two functional reaction wheels. As a consequence, fine-pointing can be maintained for approximately 6--12-hours by pointing the spacecraft solar panels toward the normal of the Sun. Once the roll motion is greater than a set threshold, a thruster is fired to reset the pointing to nominal. This results in a characteristic pattern of roll and reset every 6 hours. Fortunately, methods have been developed that effectively remove roll induced systematics \citep{Vanderburg2014}. We apply this method to these data using the kepsff code from \texttt{PyKE} \citep{Still2012}. We show the result of this in the upper panel of Figure~\ref{fig2}, where it can be seen that there is also significant variability that we attribute primarily to star-spots. We remove this stellar variability through the use of a high-pass median filter with width of 1.5 days. Our final step was to go through the light curve by hand and remove data collected during a stellar flare.

\section{Light Curve Analysis}

We modeled the K2 photometric transits of each planet in the Trappist-1 system using JKTEBOP code \citep{Southworth2008}, which employs Levenberg--Marquardt minimization to locate optimal model fits, and a residual-permutation algorithm to determine the error estimates for the derived parameters. Due to the limited quality of K2 optical photometry for this faint target ($V=18.8$), the transit shape and system parameter refinements are not goals of this work. Moreover, while variations in the transit mid-times have been observed, none of the other transit parameters for the Trappist-1 planets have been observed to vary over the course of the Spitzer observations \citep{Gillon2017}. We therefore estimated the timing series for each planet in Trappist-1 system by allowing \textit{only} the transit mid-times, $T_{\rm c}$, as well as the light-curve specific baseline flux, $F_{0}$, to float, while holding the remaining parameters fixed. We set the basic transit parameters -- the planet-to-star radius ratio, $R_{\rm P}/R_{*}$, the scaled semimajor axis, ${a}/({R_{*}+R_{\rm P}})$, the orbital inclination, $i$, and orbital period, $P$, -- to the best-fit values derived from the high-precision Spitzer photometry \citep{Gillon2017}. We hold the two quantities $e\cdot{\rm cos}\,\omega$ and $e\cdot{\rm sin}\,\omega$ relating the eccentricity, $e$, and the argument of periastron, $\omega$, (which are typically poorly constrained by a single transit observation) fixed to 0. A linear and quadratic limb-darkening law  was adopted, with coefficients fixed to the tabulated values $\mu_{1,Kepler}=0.369$ and $\mu_{2,Kepler}=0.356$ \citep{Claret2011}, using the spectroscopic stellar parameters -- effective temperature, $T_{\rm eff}=2550\,{\rm K}$, metallicity, $\rm{[Fe/H]}=0.04$, and surface gravity, ${\rm log}\,g=5.227$, -- from \citet{Gillon2017} and \citet{Viti1999}.

41, 29, 16, 11, 6, 6, and 2 transit mid-times are measured from K2 data for Trappist-1 b, c, d, e, f, g, h, respectively. Because of limited precision of K2 data for this faint target ($V=18.8$), we cannot measure the transit mid-times precisely for double or triple transit events, or transits that occur during the flares, and so all such events are omitted here. Our measured transit mid-times are shown in Figure~\ref{fig3} (black dots) with $1\,\sigma$ error bars. In order to validate our approach, we conducted transit mid-time measurements for the Trappist-1 Spitzer photometry presented by \citep{Gillon2017}, finding excellent agreement with \citet{Gillon2017}'s results.

\section{System Parameter Improvement}

\subsection{Orbital Ephemerides}
\begin{figure*}
\vspace{0cm}\hspace{0cm}
\includegraphics[scale=0.48]{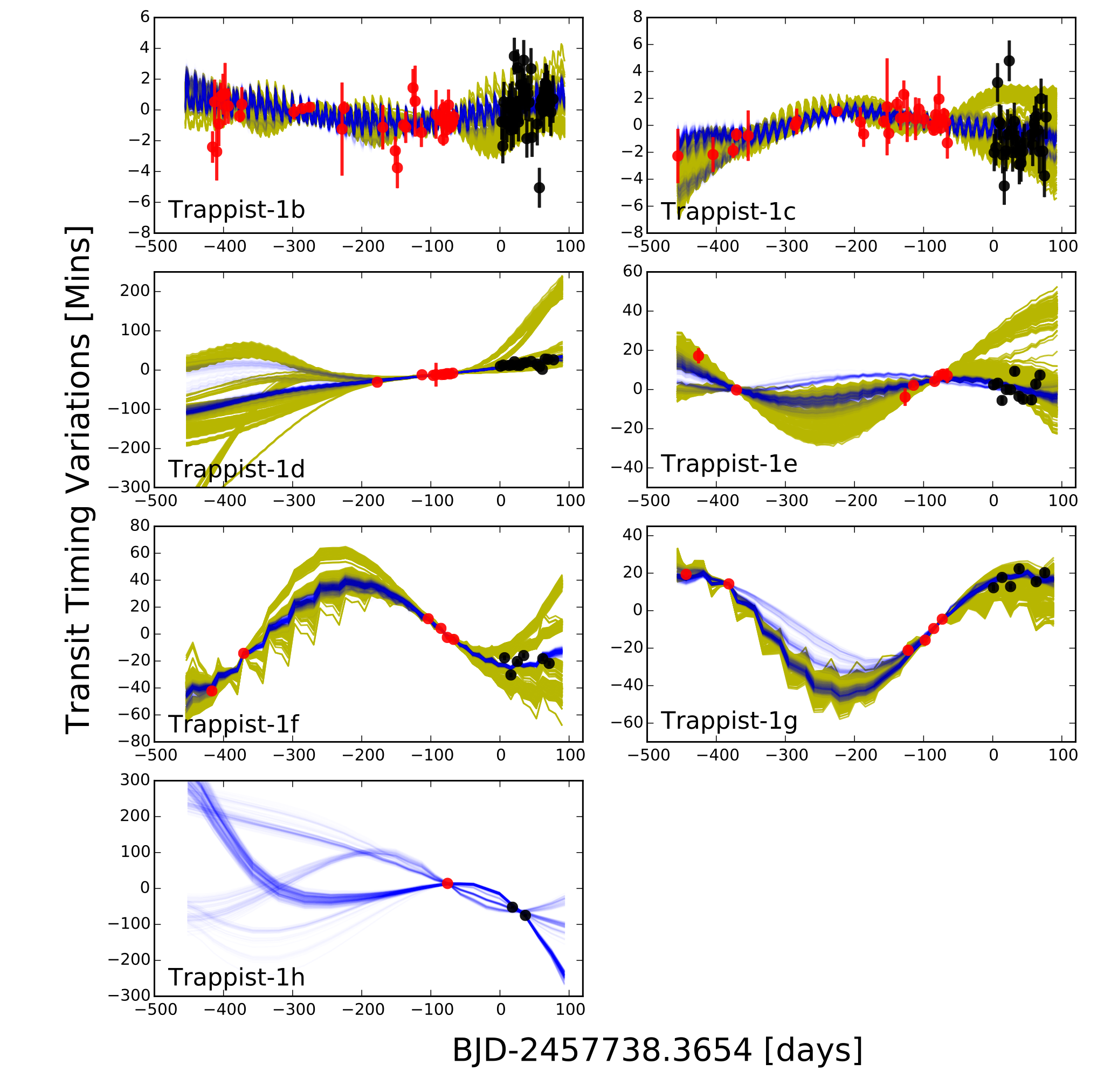}
\caption{Transit timing variations and dynamical fits for Trappist-1 planetary system. Transit timing variations from K2 photometry (black dots) and discovery data (red dots), are plotted with $1\,\sigma$ error bars. 
The lines are 1000 dynamical model drawn randomly from the converged Markov Chain in TTV dynamical fits for the discovery data + K2 photometry (blue lines) and for the discovery data (yellow lines). Extended K2 photometry permits improved mass determinations for the planets in the system.
\label{fig3}}
\end{figure*}

With the aim of refining the transit ephemerides, and thereby allowing accurate planning for observations of future occultations, we fit all the transit mid-times, $T_{\rm C}$, reported in both the \citet{Gillon2017} discovery paper and in this work as linear functions of transit epoch number ($E$),\begin{equation} \label{eq1}
T_{\rm C}[N_E]=T_{\rm C}[0]+N_E \times P\, ,
\end{equation}
with he reference epoch, $T_{\rm C}[0]$, chosen to minimize its covariance with the orbital period. To provide conservative uncertainty estimates for use in telescope scheduling, the uncertainties for the transit mid-times in the fitting process were rescaled to give a reduced $\chi^2$ equal to unity. We note, however, that the uncertainties of transit mid-times used in TTV analysis in the next section are not rescaled in this way, nor are the error bars shown in Figure~\ref{fig3}. The refined orbital ephemerides are given in Table~\ref{tab1} and show good agreement with those of \citet{Gillon2017}, except that we find slightly longer ($1.2\,\sigma$) period for Trappist-1d.   

\begin{table*}[htbp]
\begin{center}
\tiny
\caption{Updated parameters of the Trappist-1 planetary system.}
\resizebox{\textwidth}{!}{
\label{tab1}
\begin{tabular}{cccccccc}

\tableline\tableline
Star mass$M_{*}$($M_{\rm \odot}$)&\multicolumn{2}{c}{0.0802 $\pm$ 0.0073}&&&&&\\
Magnitude&\multicolumn{2}{c}{$V=18.8, R=16.6, I=14.0, J=11.4, K=10.3$}&&&&&\\\hline

Planets&b&c&d&e&f&g&h\\\hline
$T_{0}-2450000.0\,(BJD_{\rm TDB})$& $7606.56117 \pm 0.00058$ & $7568.58230 \pm 0.00064$ & $7682.2921 \pm        0.0023$ & $7574.9829 \pm 0.0025$ & $7616.1548    \pm 0.0072$ & $7529.4724 \pm    0.0058$ & $7700.0875 \pm 0.0018$\\
Period(days)& $1.5108739 \pm 0.0000075$ & $2.421818 \pm 0.000015$ & $4.04982 \pm 0.00017$ & $6.099570 \pm 0.000091$ & $9.20648 \pm 0.00053$ & $12.35281 \pm  0.00044$ & $18.76626      \pm  0.00068$ \\
Planetary Mass($\rm M_{\oplus}$) & $0.79 \pm 0.27$ & $1.63 \pm 0.63$  & $0.33 \pm 0.15$ & $0.24^{+0.56}_{-0.24}$ & $0.36 \pm 0.12$ & $0.566 \pm 0.038$ & $0.086 \pm 0.084$ \\
Semi-major Axis(AU)& 0.01111 & 0.01522 & 0.02145 & 0.02818 & 0.0371 & 0.0451 & 0.0596\\
Eccentricity& $0.019 \pm 0.008$ & $0.014 \pm 0.005$ & $0.003^{+0.004}_{-0.003} $ & $0.007 \pm 0.003$ & $0.011 \pm 0.003$ & $0.003 \pm 0.002$ & $ 0.086 \pm 0.032$ \\
Planetary Radius$^a$($R_{\rm \oplus}$)& $1.086 \pm 0.035$ & $1.056 \pm 0.035$ & $0.772 \pm 0.030$ & $0.918 \pm 0.039$ & $1.045 \pm 0.038$ & $ 1.127 \pm 0.041$ & $0.715 0.047 $ \\
Density($\rm{g/cm^{3}}$)& $3.4 \pm 1.2 $ & $7.63 \pm 3.04$ & $3.95 \pm 1.86$ & $1.71^{+4.0}_{-1.71} $ & $1.74 \pm 0.61$ & $2.18 \pm 0.28$ & $1.27 \pm 1.27$\\

Equilibrium temperature,$T_{\rm eff}$(K) & 400 & 342 & 288 & 251 & 219 & 199 & 167 \\\hline

\end{tabular}}
 \begin{flushleft}
Note. The orbits of all planets are assumed to be coplanar.\\
~~~~~a. The planetary radii are from \citet{Gillon2017} and \citet{luger2017}.
\end{flushleft}
\end{center}
\end{table*}

\subsection{TTV Analysis}
We carry out an MCMC-based dynamical fitting analysis to constrain the orbital parameters (especially the planetary masses and eccentricities) of the currently known planets in Trappist-1 system. Our analysis uses as input both the observed transit mid-times from \citet{Gillon2017} as well as those we have extracted from the newly acquired K2 data. We used the TTVFast code developed by \citet{Deck2014} to compute model transit times under the assumption that all of the planets have co-planar orbits. The 35 free parameters considered by the MCMC algorithm are the planetary masses, $M_{\rm p}$, the orbital periods, $P$, the eccentricities, $e$, the arguments of periastron, $\omega$, and the initial mean anomalies, $M_0$. We assume the same randomly distributed priors on the masses that were adopted by \citet{Gillon2017}, and we adopt random priors based on their reported period uncertainties. Our priors on the eccentricities are randomly distributed between 0 and 0.1, and those on the arguments of periastron are randomly distributed between 0 and 360 degrees. We choose 600 sets of initial values and run an independent MCMC assessment with $10^6$ iterations from each of the initial values. The first $2\times10^5$ iterations of each chain are discarded to eliminate a bias from the burn-in portion of the chain. The statistics of the parameters are derived from the last $8\times10^5$ elements of each MCMC assessment, of which every hundredth model was saved for evaluation. Our resulting planetary masses and orbital parameters, which are listed in Table~\ref{tab1}, are derived from the median and standard deviation of the $\sim 5000$ converged models with reduced $\chi^2<3.16$.

As a test of our procedure, we modeled the TTV data from the discovery paper and found results that are full consistent with those reported by \citet{Gillon2017}.

\begin{figure}
\vspace{0cm}\hspace{0cm}
\includegraphics[scale=0.255]{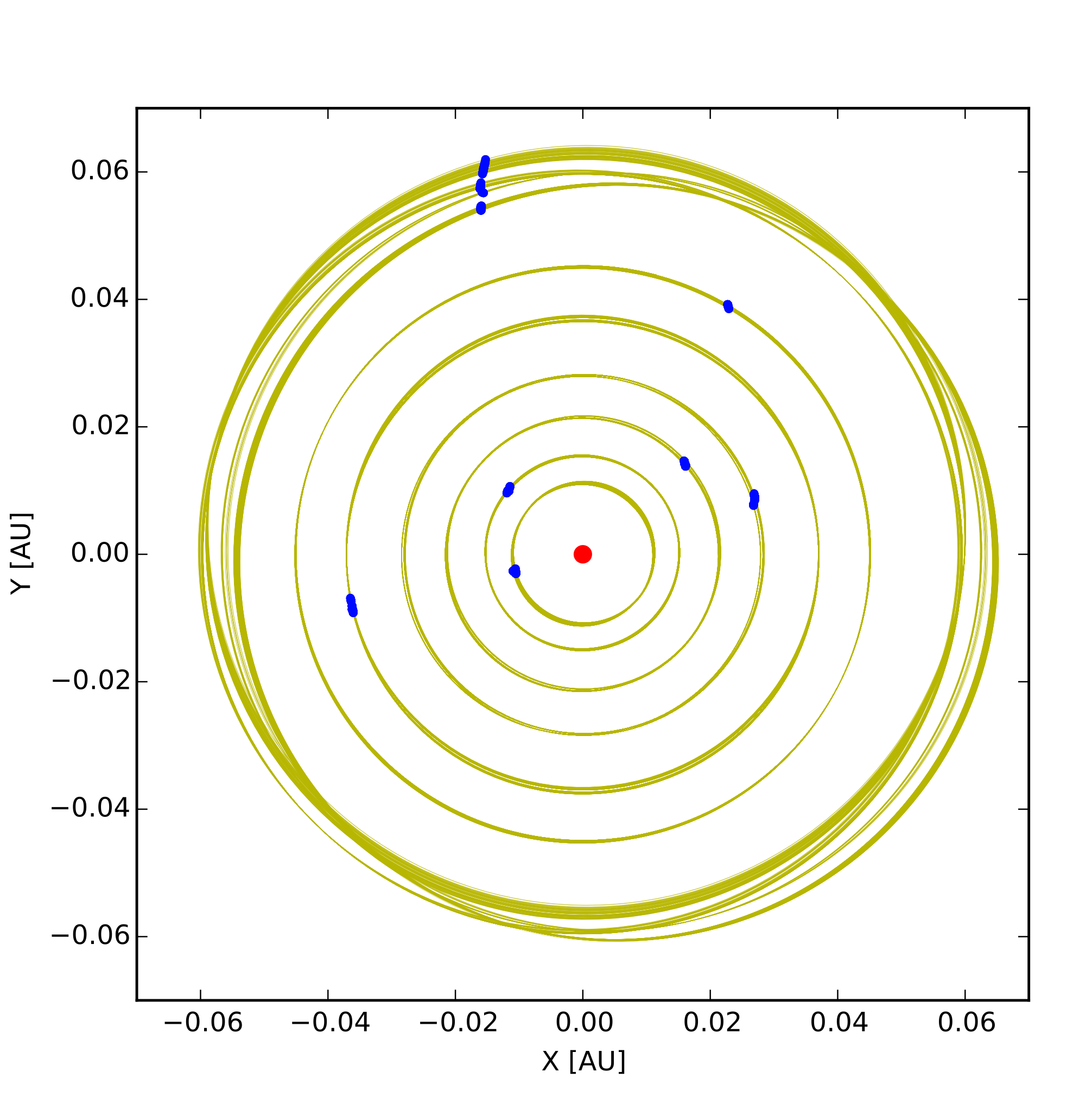}
\caption{Orbits of the Trappist-1 planetary system. 
The yellow lines are 1000 planetary orbits drawn randomly from the converged Markov Chain in TTV dynamical fits.
The blue points correspond to the location of the planets at the K2 initial epoch 2,457,738.3654.
Further photometric transit follow-up is urgently needed to attain a precise understanding of the Trappist-1 system.
\label{fig4}}
\end{figure}

\begin{figure}
\vspace{0cm}\hspace{0cm}
\includegraphics[scale=0.6]{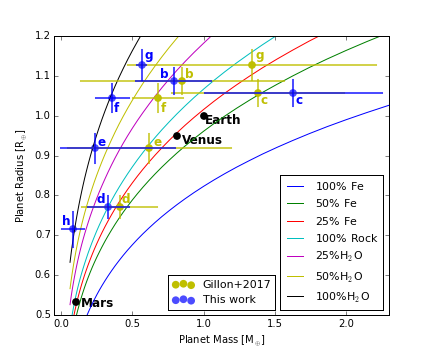}
\caption{Planetary masses and radii for Trappist-1 system. Transit timing-inferred masses from the discovery paper (yellow dots) and this work (blue dots) are plotted with $1\,\sigma$ error bars. Venus, Mars and, Earth are shown as black dots. Theoretical mass-radius relationships for different planetary compositions from \citet{Zeng2016} are plotted as colored curves.
\label{fig5}}
\end{figure}

\section{Discussion}

To within the still-substantial uncertainties, our model for the TRAPPIST-1 system is in good overall agreement with the model published by \citet{Gillon2017}. Certainly, this concordance is to be expected. The timing measurements obtained with the Spitzer spacecraft -- which anchor both sets of models -- are excellent, due in large part to the fact that the low-mass primary is substantially brighter in the infrared than in the visible ($K=10.3$ versus $V=18.8$). The extreme redness of the star permits a commensurate gain in the signal-to-noise of infrared relative to optical photometry. Nonetheless, the K2 photometry does add substantially to the overall time base line over which transits have been observed, and it cements fuller coverage of the TTV super periods, and the contribution from the K2 data is instrumental in permitting the first determination of the Mars-like mass of the outermost planet h.

Perhaps the most significant conclusion that emerges from our analysis is that the masses of the outer planets, d, e, f, and g all show noticeable decreases in comparison to the values reported by \citet{Gillon2017}. For example, the masses of planets e, f, and g (which have equilibrium temperatures of 251~K, 219~K, and 199~K, respectively) have decreased from $M_{\rm e}=0.62\,M_{\oplus}$, $M_{\rm f}=0.68\,M_{\oplus}$ and  $M_{\rm g}=1.34\,M_{\oplus}$ to $M_{\rm e}=0.24\,M_{\oplus}$, $M_{\rm f}=0.36\,M_{\oplus}$ and  $M_{\rm g}=0.57\,M_{\oplus}$. If confirmed by continued photometric monitoring of the system, such zeroth-order adjustments to the nominal masses would surely be grist for speculations concerning both the planets' physical characteristics and their possible modes of formation. Figure \ref{fig4} indicates that -- to within the errors of our determinations -- the four most distant planets are consistent with pure water compositions, and in any event, are substantially less dense either Mars or Venus. 

The lower masses and eccentricities that we have derived lead to a nominal system that shows no immediate signs of dynamical instability when integrated. This contrasts with the nominal model of \citet{Gillon2017}, for which most orbital configurations drawn from the derived parameter distributions are unstable on short time scales \citep{Tamayo2017}.

The planets orbiting TRAPPIST-1 arguably constitute the most important exoplanetary system yet discovered. The planets' large observed transit depths, coupled with the occurrence of extensive transit timing variations, present an extraordinary opportunity to discern the masses, the densities, the compositions, and the  dynamical architecture of low-mass worlds. As more data are collected, substantial insights will be gained by an evolving comparison of these these newly detected planets to the familiar terrestrial worlds of our own solar system.

\textbf{Acknowledgments}  \\
Songhu Wang gratefully acknowledges the award of a Heising-Simons 51 Pegasi b Postdoctoral Fellowship.
This material is based upon work supported by the National Aeronautics and Space Administration through the NASA Astrobiology Institute under Cooperative Agreement Notice NNH13ZDA017C issued through the Science Mission Directorate. We acknowledge support from the NASA Astrobiology Institute through a cooperative agreement between NASA Ames Research Center and Yale University.
This paper includes data collected by the \emph{K2} mission. Funding for the \emph{K2} mission is provided by the NASA Science Mission directorate.

We thank Ming Yang for his help in double-checking the transit mid-time measurements extracted from a preliminary version of the K2 light curve, and we thank Sarah Millholland for additional useful discussions.

\end{document}